\documentclass[epj,nopacs,final]{svjour}
\usepackage{graphicx}
\usepackage{amsfonts,amsmath,amssymb}
\usepackage{color}
\usepackage[colorlinks=true,urlcolor=blue,linkcolor=blue,citecolor=blue]{hyperref}
\usepackage[T1]{fontenc}
\usepackage{lmodern}
\usepackage{cite}
\begin{document}
\renewcommand{\theenumi}{(\alph{enumi})}
\title{\boldmath Searching for a hidden charm $h_1$ state in the $X(4660) \to \eta h_1$ and $X(4660) \to \eta D^* \bar D^*$ decays}
\author{Weihong Liang\inst{1,2} \and M.~Albaladejo\inst{2} \and{E.~Oset}\inst{2}
}                     
%
%
\institute{Department of Physics, Guangxi Normal University, Guilin, 541004, P. R. China \and Departamento de
F\'{\i}sica Te\'orica and IFIC, Centro Mixto Universidad de
Valencia-CSIC\\ Institutos de Investigaci\'on de Paterna, Aptdo.
22085, 46071 Valencia, Spain}
\authorrunning{W. Liang et al.}
\titlerunning{Searching for a hidden charm $h_1$ state in the $X(4660) \to \eta h_1$ and $X(4660) \to \eta D^* \bar D^*$ decays}
\date{}
%
\abstract{
We explore the possibility of experimentally detecting a predicted $h_1 ~[I^G(J^{PC})=0^-(1^{+-})]$ state of hidden charm made out from the  $D^* \bar D^*$ interaction. The method consists in measuring the decay of $X(4660)$ into $\eta D^* \bar D^*$, determining the binding energy with respect to the $D^* \bar D^*$ threshold from the shape of the $D^* \bar D^*$ invariant mass distribution. A complementary method consists in looking at the inclusive  $X(4660) \to \eta X$ decay, searching for a peak in the $X$ invariant mass distribution given by the missing $X(4660)$, $\eta$ mass. We make calculations to determine the partial decay width of $X(4660) \to \eta h_1$ from the measured $X(4660) \to \eta D^* \bar D^*$ distribution. This estimation should serve in an experiment to foresee the possibility of detecting the $h_1$ state on top of the background of inclusive events. 
%
} 
\maketitle
\section{Introduction}
The world of heavy quarks, charm and beauty, is experiencing a fast development.
A rich spectrum of new states is being found in present laboratories by the collaborations BABAR, CLEO, BELLE,
 BES \cite{Ali:2011vy,Gersabeck:2012rp,Olsen:2012zz,Li:2012pd}. The coming
 facility of FAIR will certainly add new states corresponding to quantum numbers
  which are not accessible with present machines. The states capturing more
  attention are those that do not fit within the standard picture of mesons as
  $q \bar q$ or baryons as $qqq$, and which require more complex structures,
  like tetraquarks, meson molecules, or hybrids including possible glueballs,
  for mesons, or pentaquarks and meson baryon molecules for baryons. Some of
  these states, heavy but with no open charm or beauty, would be candidates for
  quarkonium, $q \bar q$ pairs with heavy quarks, but they do not fit within the
  ordinary spectrum of such states and they have been called X,Y,Z states. The
  search for more states and theoretical work to understand their structure is a
  thriving field at present.

 The possible existence of more sophisticated states than $q \bar q$ for mesons or $qqq$ for baryons, like the multiquark states, hybrid mesons and
mesonic molecules has been early discussed within quark models \cite{Swanson:2005tq,Rosner:2006sv,Ebert:2005nc,Maiani:2005pe}. More recently, the discovery of the X,Y,Z states has stimulated much work in this direction \cite{sumrules, Ortega:2010qq, Ortega:2012rs, Branz:2009yt, Lee:2009hy, Dong:2013iqa, HidalgoDuque:2012ej}.

   One of the methods that has proved efficient to study such states is the use
   of effective field theories. Much before the X,Y,Z states were discovered,
   effective field theories were used to study the interaction of mesons among
   themselves or mesons with baryons. In many cases if was found that the
   interaction between these hadrons was attractive and strong enough to
   generate bound states or resonances, which were called dynamically generated,
   leading to some kind of molecular states of two hadrons. The use of chiral
   Lagrangians with the application of unitary techniques in coupled channels
   led to the chiral unitary approach, nowadays broadly used, which was very
   successful describing the interaction of hadrons and making predictions for
   bound states and resonances that have been verified experimentally (see ref.~\cite{review}
   for a review on this issue).

The existence of heavy meson molecules was predicted
almost 40 years ago by Voloshin and Okun \cite{Voloshin:1976ap}. In the charm sector, the field of meson molecules has been much studied
   \cite{Kolomeitsev:2003ac,Hofmann:2003je,Guo:2006fu,dany,danyax,Branz:2009yt,
   Faessler:2007gv,Segovia:2008zz,FernandezCarames:2009zz,Gutsche:2010zza, litojuan,
   HidalgoDuque:2012pq,Guo:2008zg,Ding:2009vj,Li:2012ss,hanhart,Li:2012mqa,  Albaladejo:2013aka} and many of the observed states with hidden
   charm and open charm are shown to be consistent with the molecular
   interpretation, with a good reproduction of the different observables of
   those states.

   One of the important steps in this direction was the realization that some of the X,Y,Z states could be interpreted in terms of vector-vector molecules with hidden charm \cite{raquelxyz}.  This work follows the work on the $\rho \rho$ interaction of ref.~\cite{nicmorus} using the local hidden gauge approach \cite{hidden1,hidden2,hidden4}, where the $f_2(1270)$ and $f_0(1370)$ states were interpreted as quasibound $\rho \rho$ states.\footnote{For a different interpretation of the $f_0(1370)$ as a component of an unmixed scalar octet see ref.~\cite{Albaladejo:2008qa}.} The work was extended to the SU(3) sector in ref.~\cite{gengvec} and more resonances were found, most of which could  be associated to known resonances. One of the resonances predicted in ref.~\cite{gengvec} was an $h_1$ resonance with quantum  numbers $I^G(J^{PC})=0^-(1^{+-})$ and mass around 1800 MeV, which couples to $K^* \bar K^*$. This resonance is not catalogued in the PDG \cite{pdg} and there are reasons for it since due to its negative C-parity it cannot decay into pairs of $\rho \rho$, $\omega \omega$ and equally does not couple to $\phi \phi$. For reasons of parity, coming from two vectors in $L=0$, it does not decay into a pair of pseudoscalar mesons either. This largely limits the number of decay channels, rendering the  $K^* \bar K^*$ as the channel to search for it. However, since the state lies around the $K^* \bar K^*$ threshold, a neat peak might be difficult to see and would be distorted by the limited phase space around threshold. Nevertheless, it was very recently found \cite{migueh1} that a peak around the $K^* \bar K^*$ threshold, seen in the   $J/\psi \to \eta K^{*0}\bar{K}^{*0}$ reaction observed at BES \cite{BESdata}, was naturally interpreted as a signature of the $h_1$ resonance predicted in ref.~\cite{gengvec}.

   One can guess that the $h_1$ state made out of $K^* \bar K^*$ could have an analogue in the $D^* \bar D^*$, and indeed this is the case as found in ref.~\cite{raquelxyz}. There such a state was found around 3945 MeV and couples mostly to $D^* \bar D^*$ and less strongly to $D^*_s \bar D^*_s$. The same state, with mass 3955 $\pm$ 16 MeV, is obtained using heavy quark spin symmetry in refs.~\cite{HidalgoDuque:2012pq, litojuan}.  For the same reasons as before, this state does not couple to any other pair of vector mesons, except $K^* \bar K^*$, but with such a tiny coupling that makes fruitless its investigation in this channel. Once again we have to resort to the $D^* \bar D^*$ channel, with the added problem that now it is a bound state in this channel and the $D^*$ has a very small width.  Yet, as we shall see, it is possible to find a signature of the resonance by looking at the $D^* \bar D^*$ spectrum in the $X(4660) \to \eta h_1$ and $X(4660) \to \eta D^* \bar D^*$ reactions. The choice of the $X(4660)$ is made in order to find a similar reaction as the one of ref.~\cite{BESdata}, replacing the $J/\psi$ by the only analogous vector which has enough energy to produce the reaction. The $X(4660)$ is catalogued in the PDG as a $? ^?(1^{--})$ state.  Yet, the G-parity should be negative since it decays into $\psi(2S) \pi^+ \pi^-$. And the theoretical papers written on this state also attribute to it zero isospin \cite{polosa,Shi:2013zn,Guo:2010tk}.

   Another way to find the predicted resonance would be to search in the inclusive $X(4660) \to \eta X$ reaction, where $X$ is undetermined,\footnote{We denote with $X$ whatever undetermined state appears as a decay product together with the $\eta$ meson. To avoid confusion with the state $X(4660)$, this one will always appear in this work with its nominal mass between parenthesis.} but looking at the $X$ invariant mass distribution determined from the observation of the $\eta$ alone. This might not be easy if one has a large background of $X(4660) \to \eta X$ events, since one must look for a probably small signal over a relatively large background. It is then most convenient to know the rate expected before planning the experiment. This is one additional information we provide here. We show that the knowledge of the $X(4660) \to  \eta D^* \bar D^*$ partial decay width allows us to determine the partial decay width for $X(4660) \to  \eta h_1$. The rates obtained for these two decay modes are of comparable size, and the strength of the $X(4660) \to  \eta h_1$ is concentrated in a narrow window of invariant masses of $h_1$, its small width. Thus, provided that the integrated $X(4660) \to  \eta D^* \bar D^*$ width is of the order of the background of the $X(4660) \to \eta X$ reaction, the probability that a neat peak in the $X(4660) \to  \eta X$ inclusive reaction can be seen is large. The theoretical study of the $X(4660) \to  \eta D^* \bar D^*$ and $X(4660) \to  \eta h_1$ reactions is the topic of the present paper, with the purpose of providing information that motivates the experimental search of this otherwise elusive resonance.

\section{Formalism}

\begin{figure}[t!]\centering
\includegraphics[height=2.3cm,keepaspectratio]{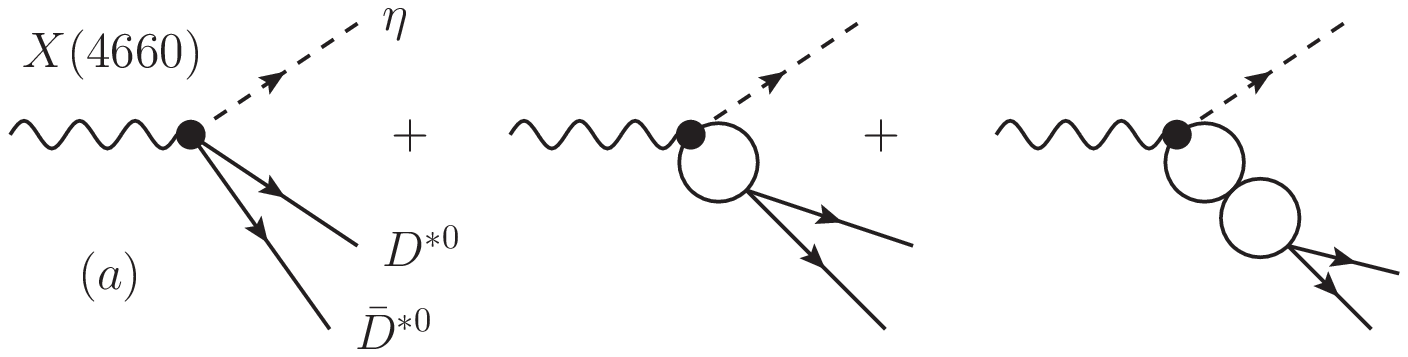}
\includegraphics[height=2.3cm,keepaspectratio]{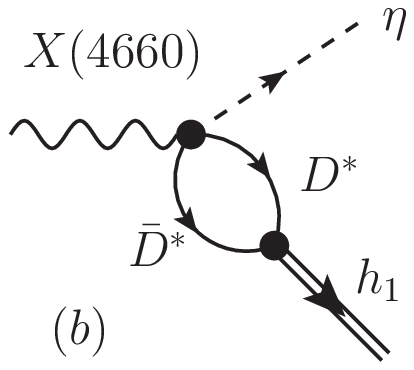}
\caption{Decay mechanism of the $X(4660)$ resonance. In $(a)$, the three-body decay $X(4660)\to \eta D^{\ast 0}\bar{D}^{\ast 0}$ is shown. In $(b)$ we represent the decay $X(4660) \to \eta h_1$, where the $h_1$ is generated by the $D^{\ast}\bar{D}^{\ast}$ interaction.\label{fig:diag1}}
\end{figure}

In studying the $X(4660) \to \eta h_1$ and $X(4660) \to \eta D^{\ast 0} \bar{D}^{\ast 0}$ decay processes, we shall follow the formalism of ref.~\cite{migueh1}, which closely follows the idea of refs.~\cite{ollerpsi, chiang} in the $J/\psi \to \phi f_0(980)$ decay. The formalism is suited to study the decay of a particle into a spectator and a dynamically generated resonance. Since the latter is produced via the interaction of a pair of particles, the mechanism consists of a primary decay into the spectator and the pair of particles (without interaction), followed by the interaction of the pair to generate the resonance. For the particular case that we study here, the corresponding Feynman diagrams are depicted in fig.~\ref{fig:diag1}. However, we shall differentiate two situations. In the first one, the $I=0$ $D^{*0} \bar D^{*0}$ pair is explicitly present in the final state [fig.~\ref{fig:diag1}(a)]. In the second one, the $\eta$ and the $h_1$ resonance are produced, regardless of which its decay channel is [fig.~\ref{fig:diag1}(b)]. This situation is faced when the inclusive $X(4660) \to \eta X$ decay is studied, and one looks for a peak in the square of the invariant mass distribution ${(P_{X(4660)}-p_{\eta})}^2$.

The $h_1$ which is the subject of this work is dynamically generated in the $D^\ast \bar{D}^\ast$ interaction in $I=0$. To write the amplitude $t$ for this process, we shall make use of the on-shell factorized form of the Bethe-Salpeter equation \cite{ollerulf},
\begin{align}\label{eq:tmat}
t(s) & = v(s) + v(s) G(s) t(s) \nonumber \\
& = v(1+Gt) = \left( v^{-1} - G \right)^{-1}~,
\end{align}
where $s\equiv M_\text{inv}^2$ is the invariant mass squared of the $D^\ast \bar{D}^\ast$ system. The function $v$ in eq.~\eqref{eq:tmat} is the potential for the $D^\ast \bar{D}^\ast$ interaction in $I=0$ and $L=0$. Two different forms of the potential will be used in this work. The first one is the dynamical potential of ref.~\cite{raquelxyz}, and the second one is a constant potential. We will discuss both approaches below. In eq.~\eqref{eq:tmat}, $G$ is the one-loop two-point function for the $D^\ast \bar{D}^\ast$ system, conveniently regularized. In this work, we use a once-subtracted form for this function. For the case of equal masses running through the loop, $G$ can be written as
\begin{equation}\label{eq:Gfun}
G(s) = \frac{1}{16\pi^2} \left( a(\mu) + \log \frac{m_{D^\ast}^2}{\mu^2} - \sigma \log \frac{\sigma - 1}{\sigma + 1} \right)~,
\end{equation}
with $\sigma(s) = \sqrt{1 - 4m_{D^\ast}^2/s}$. The parameter $a(\mu)$ is the subtraction constant, and it depends on the regularization scale $\mu$ such that $a(\mu) - \log \mu^2$ is independent of $\mu$, that is, there is only one free parameter. The presence of the function $G$ in eq.~\eqref{eq:tmat} ensures the elastic unitarity of the amplitude $t$ above threshold, $\sqrt{s} > 2m_{D^\ast}$,
\begin{equation}\label{eq:unitarity}
{\rm Im}\ t^{-1}(s) = - {\rm Im}\ G(s) = \frac{\sigma(s)}{16\pi}~.
\end{equation}

The state $h_1$ appears as an $I=0$ bound state (with mass $M_{h_1} \equiv \sqrt{s_{h_1}} < 2 m_{D^\ast}$) of the $D^\ast \bar{D}^\ast$ system, that is, as a pole in the physical Riemann sheet of the amplitude in eq.~\eqref{eq:tmat},
\begin{equation}\label{eq:tmat_pole}
t = \frac{g^2}{s-s_{h_1}}~,
\end{equation}
where $g$ is the coupling of the state $h_1$ to the $D^\ast \bar{D}^\ast$ system. The quantity $gG$ is related to the wave function around the origin \cite{danyjuan}. The coupling $g$ can be calculated as the residue of the amplitude at the pole $s=s_{h_1}$. However, from eq.~\eqref{eq:tmat_pole}, it can be also calculated as:
\begin{equation}\label{eq:g2cou}
\frac{1}{g^2} = \frac{d t^{-1}}{d s} = \frac{d v^{-1}}{d s} - \frac{d G}{d s}~,
\end{equation}
where the derivative is obviously evaluated at $s=s_{h_1}$. In the case of small binding energies $B=2m_{D^\ast} - M_{h_1}$, this derivative is driven by the unitarity term in eq.~\eqref{eq:unitarity} analytically extrapolated below threshold, and it can be simplified to give:
\begin{equation}\label{eq:g2cou_wei}
g^2 = 64\pi m_{D^\ast} \sqrt{4B m_{D^\ast}}~,
\end{equation}
which is nothing but Weinberg's formula for the coupling of a weakly bound state \cite{weinberg, hanhartwei}.

Now we discuss about the potential $v$ for the $D^\ast \bar{D}^\ast$ interaction in $I=0$ appearing in eq.~\eqref{eq:tmat}. This potential can be calculated within the hidden gauge formalism, as in ref.~\cite{raquelxyz}, where the $h_1$ state we are studying was predicted. The expression for this potential is given by:
\begin{equation}\label{eq:dynpot}
v = \left(9 + b\left(1 - \frac{3s}{4m_{D^\ast}^2} \right)  \right) g_{D^\ast}^2 
\end{equation}
with $g_{D^\ast} = m_{D^\ast}/{2f_{D^\ast}}$, $f_{D^\ast} \simeq 146\ \text{MeV}$, and where the constant $b$ is given in terms of the masses of the vector mesons, having a value $b=27.6$. Since the potential is fixed in this case, the only free parameter of the amplitude $t$ is the subtraction constant $a(\mu)$. Hence, once this constant is fixed, the amplitude is completely determined. In particular, the position of the bound state is also fixed. This argument can be also reversed and for a given position of the bound state there corresponds a unique value of the subtraction constant $a(\mu)$.

We will also make use of a constant (energy independent) potential. This is a good approximation for the limited range of energies $M_\text{inv}$ that will be studied in this work. The use of such a potential also provides some model independence to our calculation. Besides its simplicity, a constant potential has other advantages, as we discuss now. If there is a bound state, the potential is such that
\begin{equation}\label{eq:constantpotential}
G(s_{h_1}) = v^{-1} \equiv G_{h_1}~,
\end{equation}
so that the amplitude in eq.~\eqref{eq:tmat} can be simplified to:
\begin{equation}\label{eq:tmat_cp}
t = \frac{1}{G_{h_1} - G}~.
\end{equation}
It is worth noting that the difference in the denominator does not depend on the subtraction constant $a(\mu)$ used in the loop function $G$, eq.~\eqref{eq:Gfun}.\footnote{If the loop function is regularized by means of a cutoff $\Lambda$, then this difference is independent of the cutoff in the $\Lambda \to \infty$ limit.} In this way, the amplitude is fully determined by means of just one parameter, the mass of the bound state. Furthermore, also the coupling $g$ is entirely determined by the mass of the bound state, since the derivative of the inverse of the potential in eq.~\eqref{eq:g2cou} disappears. The coupling $g$ is then given solely by the derivative of the $G$ function, which is independent of the subtraction constant. In the end, this means that the whole amplitude is determined by the bound state mass and unitarity.

We now discuss the decay processes depicted in fig.~\ref{fig:diag1}. Let us start with the three-body decay process of fig.~\ref{fig:diag1}(a). We shall assume that the primary production vertex for $X(4660) \to \eta D^\ast \bar{D}^\ast$, with the $D^\ast \bar{D}^\ast$ pair in $I=0$, is of short-range nature, that is, a constant in the field theory formalism. We denote by $V_P$ this bare vertex. The full amplitude $\widetilde{T}_P$ for that process must take into account the $D^\ast \bar{D}^\ast$ final state interaction, as shown in fig.~\ref{fig:diag1}(a). The full amplitude is then:
\begin{equation}
\widetilde{T}_P = \frac{V_P}{\sqrt{2}}  \left( 1 + v G + \cdots \right) = \frac{V_P}{\sqrt{2}}  \left( 1 + v G t \right) = \frac{V_P}{\sqrt{2}} \frac{t}{v}~,\label{eq:tprelation}
\end{equation}
where eq.~\eqref{eq:tmat} has been taken into account. The factor $1/\sqrt{2}$ appears because of the $D^{\ast 0} \bar{D}^{\ast 0}$ component of the $I=0$ $D^\ast\bar{D}^\ast$ system. Making use of eq.~\eqref{eq:tprelation}, the differential decay width for the process $X(4660) \to \eta D^{\ast 0} \bar{D}^{\ast 0}$ can be written as:
\begin{align}
\frac{d \Gamma_{\eta D^{\ast 0} \bar{D}^{\ast 0}}}{d M_\text{inv}} & = \frac{\lvert \widetilde{T}_P\rvert^2}{32\pi^3} \frac{p_\eta \widetilde{p}_{D^{\ast}}}{M^2_{X(4660)}} \nonumber \\ & = \frac{\left\lvert V_P \right\rvert^2}{64\pi^3} \frac{p_\eta \widetilde{p}_{D^{\ast}}}{M^2_{X(4660)}} \left\lvert \frac{t}{v} \right\rvert^2~,\label{eq:diffwidth}
\end{align}
where $p_{\eta}$ is the $\eta$ momentum in the rest frame of the $X(4660)$ decaying into an $\eta$ and a $D^{*0}\bar D^{*0}$ pair with invariant mass $M_\text{inv}$,
\begin{equation}\label{eq:peta}
p_{\eta}=\frac{\lambda^{1/2}(M^2_{X(4660)}, m_{\eta}^2, M_\text{inv}^2)}{2 M_{X(4660)}}~,
\end{equation}
and $\widetilde{p}_{D^{*0}}$ is the $D^{*0}$ momentum in the rest frame of the $D^{*0}\bar D^{*0}$ system,
\begin{equation}\label{eq:pds}
  \widetilde{p}_{D^{*0}}=\frac{\lambda^{1/2}(M_\text{inv}^2, M_{D^{*0}}^2, M_{D^{*0}}^2)}{2M_\text{inv}}.
\end{equation}
In eqs.~\eqref{eq:peta} and \eqref{eq:pds}, $\lambda(x,y,z)$ is the K\"ahlen or triangle function, $\lambda(x,y,z) = x^2 + y^2 + z^2 -2xy-2yz-2zx$.

We now discuss the $X(4660) \to \eta h_1$ two-body decay process, for which the the amplitude $T_P$ can be obtained from fig.~\ref{fig:diag1}(b),
\begin{equation}
T_P = V_P G_{h_1} g~,
\end{equation}
where $G_{h_1} \equiv G(M_{h_1}^2)$ is the loop function calculated at the mass of the $h_1$ state. Then, the decay width for this process, $\Gamma_{\eta h_1}$, can be calculated as:
\begin{equation}\label{eq:twobodywidth}
\Gamma_{\eta h_1} = \frac{\left\lvert T_P \right\rvert^2 }{8\pi} \frac{p_{\eta,h_1}}{M^2_{X(4660)}} = \frac{\left\lvert V_P \right\rvert^2}{8\pi} \frac{p_{\eta,h_1}}{M^2_{X(4660)}}\ G_{h_1}^2\ g^2~,
\end{equation}
where $p_{\eta,h_1}$ is $p_\eta$ in eq.~\eqref{eq:peta} calculated for $M_\text{inv} = M_{h_1}$. We can now divide eqs.~\eqref{eq:diffwidth} and \eqref{eq:twobodywidth} to get rid of the unknown vertex $V_P$, and express the differential decay width of eq.~\eqref{eq:diffwidth} in terms of known quantities,
\begin{equation}\label{eq:diffwidth2}
\frac{d \Gamma_{\eta D^{\ast 0} \bar{D}^{\ast 0}}}{d M_\text{inv}}  = 
\frac{\Gamma_{\eta h_1}}{8\pi^2 g^2} \frac{p_\eta \widetilde{p}_{D^{\ast 0}}}{p_{\eta,h_1}} \left\lvert \frac{t}{v \ G_{h_1}} \right\rvert^2~.
\end{equation}
This is the final expression for the differential decay width when one uses a dynamical potential, as in eq.~\eqref{eq:dynpot}. Under the assumption of a constant potential, however, this expression can be further simplified by means of eqs.~\eqref{eq:constantpotential} and \eqref{eq:tmat_cp} to:
\begin{equation}\label{eq:diffwidth_cp}
\frac{d \Gamma_{\eta D^{\ast 0} \bar{D}^{\ast 0}}}{d M_\text{inv}}  = 
\frac{\Gamma_{\eta h_1}}{8\pi^2 g^2} \frac{p_\eta \widetilde{p}_{D^{\ast 0}}}{p_{\eta,h_1}} \left\lvert \frac{1}{G_{h_1} - G} \right\rvert^2~.
\end{equation}
It is worth stressing that in this case, the differential decay width depends only on the mass of the bound state (up to the unknown width $\Gamma_{\eta h_1}$). Another useful quantity that we can calculate is the following ratio,
\begin{equation}\label{eq:ratio1}\displaystyle
R = \frac{\Gamma_{\eta h_1}}{\Gamma_{\eta D^{\ast 0} \bar{D}^{\ast 0}}}~,
\end{equation}
where $\Gamma_{\eta D^{\ast 0} \bar{D}^{\ast 0}}$ is the integrated differential decay width in eq.~\eqref{eq:diffwidth2},
\begin{equation}\label{eq:ratio2}
\Gamma_{\eta D^{\ast 0} \bar{D}^{\ast 0}} = \int{d M_\text{inv} \frac{d \Gamma_{\eta D^{\ast 0} \bar{D}^{\ast 0}}}{d M_\text{inv}}}~,
\end{equation}
where the integration runs over $2m_D^\ast < M_\text{inv} < M_{X(4660)} - m_\eta$, the available range of energies for the $D^{\ast 0} \bar{D}^{\ast 0}$ system.

\section{Results}

\begin{figure}\centering
\includegraphics[width=0.49\textwidth,keepaspectratio]{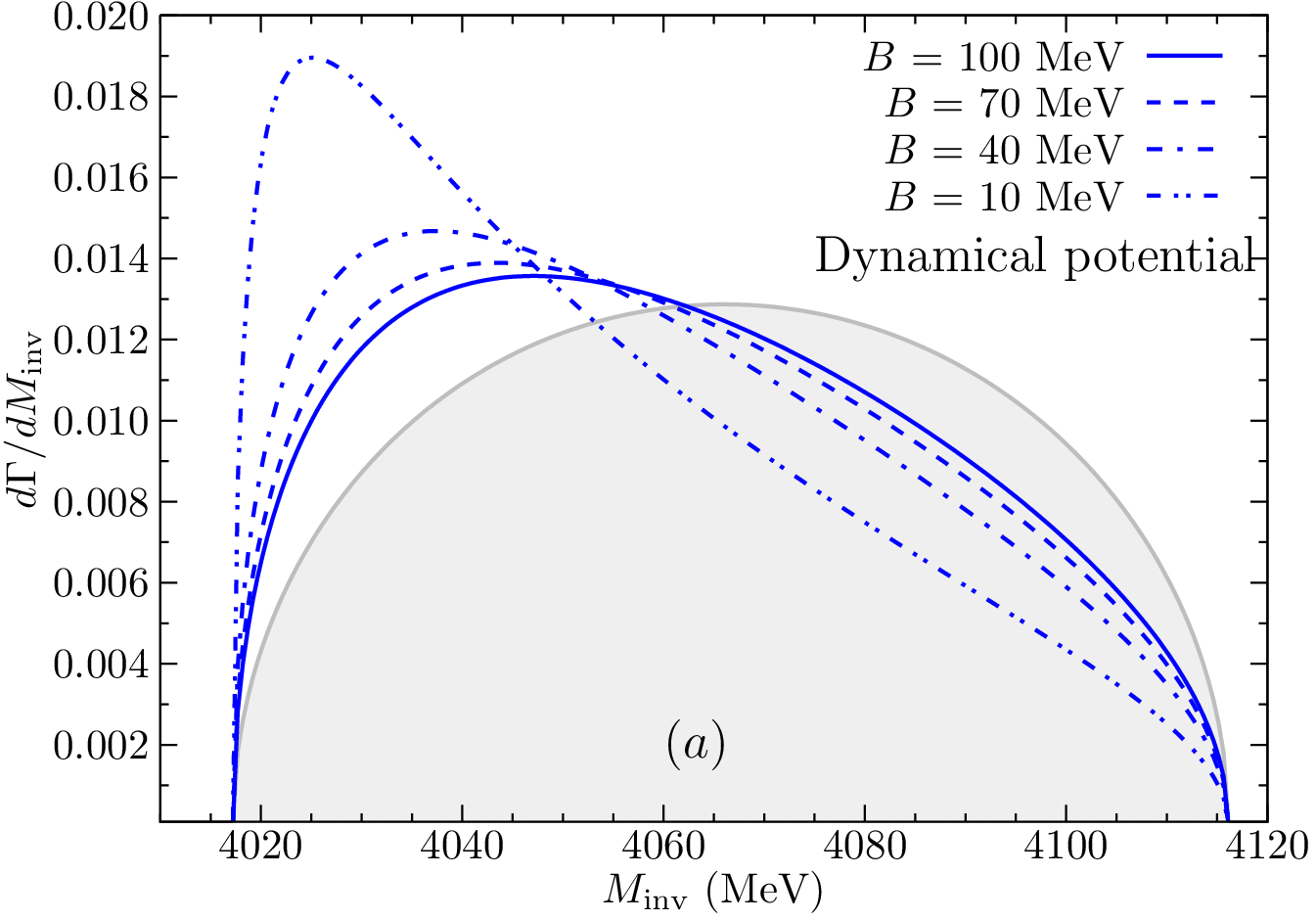}
\includegraphics[width=0.49\textwidth,keepaspectratio]{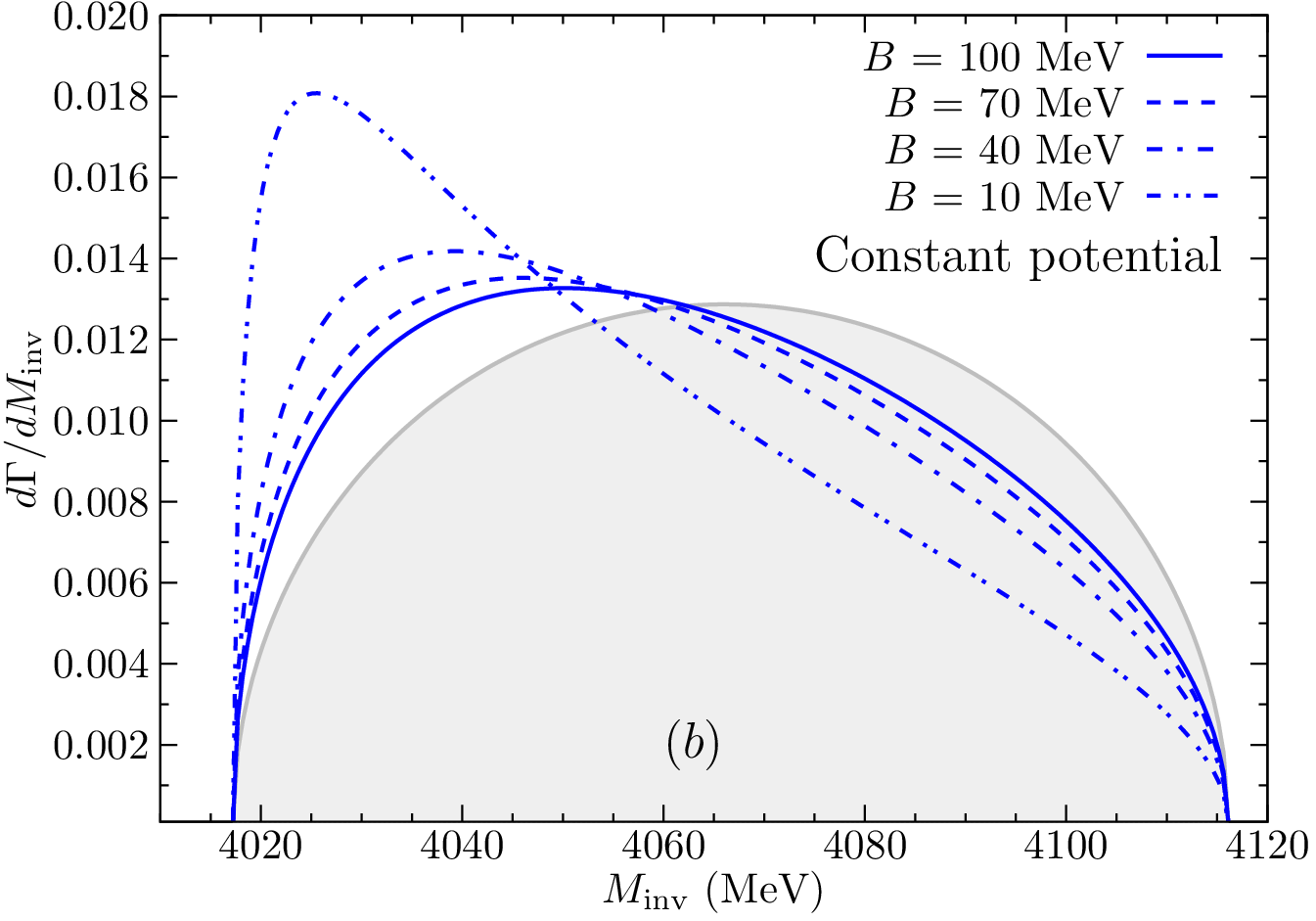}
\caption{(Color online) Differential decay width for the process $X(4660)\to \eta D^{\ast 0} \bar{D}^{\ast 0}$.  The solid, dashed, dot-dashed, and double-dot-dashed lines correspond to binding energies $B=100$, $70$, $40$, and $10\ \text{MeV}$, respectively. The shaded area stands for a phase space distribution. All curves are normalized so has to have the same area. (a) Results obtained with the dynamical potential, eqs.~\eqref{eq:dynpot} and \eqref{eq:diffwidth2}. (b) Results obtained with the constant potential, eqs.~\eqref{eq:constantpotential} and \eqref{eq:diffwidth_cp}.\label{fig:res1}}
\end{figure}

In fig.~\ref{fig:res1} we show the $X(4660) \to \eta D^{\ast 0}\bar{D}^{\ast 0}$ differential decay width in terms of $M_\text{inv}$ for the four binding energies $B=100$, $70$, $40$, and $10\ \text{MeV}$, with solid, dashed, dot-dashed, and double-dot-dashed lines, respectively. In fig.~\ref{fig:res1}(a) results obtained with the dynamical potential approach [eqs.~\eqref{eq:dynpot} and \eqref{eq:diffwidth2}] are shown. Note that if there is a bound state, there is a one-to-one correspondence between the binding energy $B$ and the subtraction constant $a(\mu)$. For the four cases represented in fig.~\ref{fig:res1}(a), the subtraction constant is given in Table~\ref{tab:usedparams}. Analogously, the results of fig.~\ref{fig:res1}(b) stem from using a constant potential, eqs.~\eqref{eq:constantpotential} and \eqref{eq:diffwidth_cp}. In both cases, the curves are compared with a phase space distribution, proportional to $p_\eta \widetilde{p}_{D^\ast}$ [eqs.~\eqref{eq:peta} and \eqref{eq:pds}]. In order to have a meaningful comparison, all the curves are normalized so that the area below them is the same. As we can see, even for large binding energies of $h_1$ the shape of the mass distribution differs substantially from phase space. As the binding energy decreases, the differences become more significant. It is clear that with reasonable statistics the shape can be well determined and the binding energy can be obtained from there, by using any of the methods presented in this work.

\begin{figure}[t!]\centering
\includegraphics[width=0.47\textwidth,keepaspectratio]{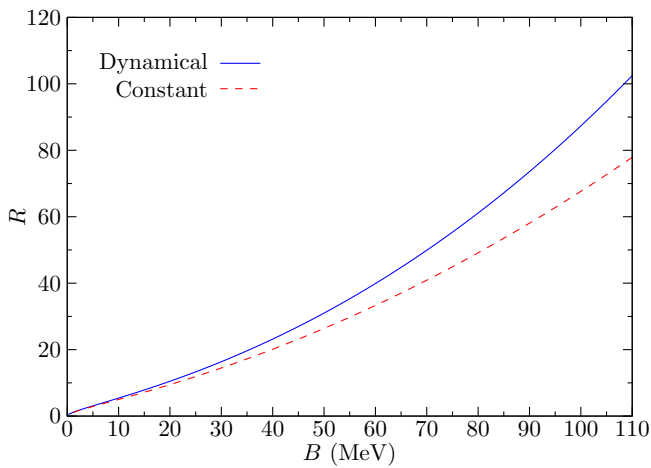}
\caption{(Color online) The ratio $R$ [eqs.~\eqref{eq:ratio1} and \eqref{eq:ratio2}] as a function of the binding energy $B$ of the $h_1$ state. The blue solid line shows the results for the dynamical potential approach, eqs.~\eqref{eq:dynpot} and \eqref{eq:diffwidth2}. The red dashed line, in contrast, is obtained with a constant potential, eqs.~\eqref{eq:constantpotential} and \eqref{eq:diffwidth_cp}.\label{fig:res2}}
\end{figure}

\begin{table}[t!]\centering
\caption{Coupling of the $h_1$ state to $D^\ast \bar{D}^\ast$. For each binding energy (first column) used in fig.~\ref{fig:res1} to calculate the differential decay width of the $X(4660)\to \eta D^{\ast 0} \bar{D}^{\ast 0}$ process, we give the coupling as calculated with eq.~\eqref{eq:g2cou} for a dynamical potential, eq.~\eqref{eq:dynpot} (second column), or with a constant potential, eq.~\eqref{eq:constantpotential} (third column). We also show the coupling as calculated with Weinberg's formula, eq.~\eqref{eq:g2cou_wei} (fourth column). The subtraction constant $a(\mu)$ (for $\mu = 1\ \text{GeV}$) needed in the loop function $G$ when the dynamical potential is used to reproduce the binding energy is shown in brackets in the second column.\label{tab:usedparams}}
\begin{tabular}{cccc}
\hline\noalign{\smallskip}
$B$ (MeV)        & $g$ (GeV) [$a(\mu)$]    & $g$ (GeV)  & $g$ (GeV) \\ 
                 & (Dyn.)                  &  (Cons.)   &  (Weinberg)  \\ 
\noalign{\smallskip}\hline\noalign{\smallskip}
           $100$ & $21.2\quad [-2.1]$      & $21.5$     & $19.0$   \\
$\hphantom{1}70$ & $19.1\quad [-2.0]$      & $19.3$     & $17.4$   \\
$\hphantom{1}40$ & $16.3\quad [-1.9]$      & $16.4$     & $15.1$   \\
$\hphantom{1}10$ & $11.1\quad [-1.7]$      & $11.2$     & $10.7$   \\
\noalign{\smallskip}\hline
\end{tabular}
\end{table}

In fig.~\ref{fig:res2} we plot the value of $R$ [eq.~\eqref{eq:ratio1}] as a function of the binding energy. As we can see, the values of $R$ range from 5 to 90 within the range of the considered binding energies. It is interesting to see that the values of $R$ are relatively large. This means that the width of $\Gamma_{\eta h_1}$ is of the same order or larger than $\Gamma_{\eta D^{\ast 0}\bar{D}^{\ast 0}}$, the integrated differential decay width. However, when looking in the inclusive reaction $X(4660) \to \eta X$, $\Gamma_{\eta h_1}$ will be distributed in a small range of the squared invariant mass of $X$, $M_\text{inc}^2 = (P_{X(4660)}-p_{\eta})^2$, which, given the small width of the $h_1$, will come mostly from the experimental resolution. 

Summarizing how the results found in this work could be experimentally used, the strategy to find the elusive $h_1$ resonance is twofold:
\begin{enumerate}
\item Measure $d \Gamma_{\eta D^{\ast 0}\bar{D}^{\ast 0}}/d M_\text{inv}$ and, from the line shape, determine the binding energy of $h_1$, according to fig.~\ref{fig:res1}.

\item By integrating $d \Gamma_{\eta D^{\ast 0}\bar{D}^{\ast 0}}/d M_\text{inv}$, and using the theoretical ratio $R$ [eqs.~\eqref{eq:ratio1} and \eqref{eq:ratio2}] of fig.~\ref{fig:res2}, determine $\Gamma_{\eta h_1}$. By measuring $d \Gamma_{\eta X}/d M_\text{inc}$ in the inclusive $X(4660) \to \eta X$ reaction, one would get a size of the background in this inclusive reaction. If $\Gamma_{\eta h_1}$ is not very small compared with this background, it would be then possible to observe a peak associated to $h_1$ on top of this background. The bonus of this second part is that the $h_1$ will now appear as a narrow peak, allowing to determine its mass and width (unless the experimental resolution is bigger than the width of the $h_1$), while in the method (a) the mass would be only indirectly obtained and no information on the width would be provided.
\end{enumerate}

In case the $h_1$ resonance width is smaller than the experimental resolution, a bound in the width can be provided. Still, provided the $h_1$ peak is visible on top of the background, the mass could be well determined, and the consistency with the mass found with method (a) could be tested. In this case, the mass determined with method (b) would be more precise than the one determined with method (a).

\section{Conclusions}
In this work we have studied the $X(4660) \to \eta D^* \bar D^*$ and $X(4660) \to \eta h_1$ reactions, where $h_1$ is an axial vector state [$0^-(1^{+-})$] which is theoretically predicted as a bound state from the $D^* \bar D^*$ interaction. The $h_1$ states made out from vector-vector interaction are very elusive since they do not decay into other pairs of vectors than those from which they are built. Also for reasons of parity, they do not decay into pairs of pseudoscalar mesons. This could justify why the $h_1$ state predicted has not yet been detected, thus the relevance of finding suitable reactions where they can be found. 

In the present work we show that a measurement of the $D^* \bar D^*$ invariant mass distribution in the $X(4660) \to \eta D^* \bar D^*$ reaction allows one to determine the $h_1$ mass. 

  On the other hand, we also explore a complementary method of analysis by looking for a peak in the $X$ mass distribution in the inclusive $X(4660) \to \eta X$ reaction. We show that one can determine the rate for the $X(4660) \to \eta h_1$ decay from the integrated spectrum of the $X(4660) \to \eta D^* \bar D^*$ reaction. This knowledge, and a comparison with the background of the inclusive $X(4660) \to \eta X$ reaction, will show the chances that one has to detect a neat peak for the $h_1$ on top of this background. This latter procedure allows one to determine with precision the mass and width of the $h_1$ state, or at least a bound for the width if the experimental resolution exceeds the value of the width.
  
    The interest of the hadron community in finding meson states that do not fit
the standard $q \bar q$ structure, together with the information provided in the present work, should stimulate the implementation of the reactions suggested which can be carried out at some present and future facilities.

\section*{Acknowledgments}

This work is partly supported by the Spanish Ministerio de Economia
y Competitividad and European FEDER funds under the contract number
FIS2011-28853-C02-01 and FIS2011-28853-C02-02, and the Generalitat
Valenciana in the program Prometeo, 2009/090. We acknowledge the
support of the European Community-Research Infrastructure
Integrating Activity Study of Strongly Interacting Matter (acronym
HadronPhysics3, Grant Agreement n. 283286) under the Seventh
Framework Programme of EU. This work is also partly supported by the
National Natural Science Foundation of China under Grant No. 11165005
and by scientific research fund (201203YB017) of education department of Guangxi.

\bibliographystyle{plain}

\begin{thebibliography}{999}

\bibitem{Ali:2011vy}
  A.~Ali,
  PoS BEAUTY {\bf 2011}, 002 (2011).

\bibitem{Gersabeck:2012rp}
  M.~Gersabeck,
  Mod.\ Phys.\ Lett.\ A {\bf 27}, 1230026 (2012).

\bibitem{Olsen:2012zz}
  S.~L.~Olsen,
  Prog.\ Theor.\ Phys.\ Suppl.\  {\bf 193}, 38 (2012).

\bibitem{Li:2012pd}
  L.~Li [BESIII Collaboration],
  Nucl.\ Phys.\ Proc.\ Suppl.\  {\bf 225-227} (2012) 107.

\bibitem{Swanson:2005tq}
  E.~Swanson,
  Int.\ J.\ Mod.\ Phys.\ A {\bf 21}, 733 (2006)
  [AIP Conf.\ Proc.\  {\bf 814}, 203 (2006)].

\bibitem{Rosner:2006sv}
  J.~L.~Rosner,
  J.\ Phys.\ Conf.\ Ser.\  {\bf 69}, 012002 (2007).

\bibitem{Ebert:2005nc}
  D.~Ebert, R.~N.~Faustov and V.~O.~Galkin,
  Phys.\ Lett.\ B {\bf 634}, 214 (2006).

\bibitem{Maiani:2005pe}
  L.~Maiani, V.~Riquer, F.~Piccinini and A.~D.~Polosa,
  Phys.\ Rev.\ D {\bf 72}, 031502 (2005).


\bibitem{sumrules}
  M.~Nielsen, F.~S.~Navarra and S.~H.~Lee,
  Phys.\ Rept.\  {\bf 497}, 41 (2010).

\bibitem{Ortega:2010qq}
  P.~G.~Ortega, J.~Segovia, D.~R.~Entem and F.~Fernandez,
  Phys.\ Rev.\ D {\bf 81}, 054023 (2010).


\bibitem{Ortega:2012rs}
  P.~G.~Ortega, D.~R.~Entem and F.~Fernandez,
  J.\ Phys.\ G {\bf 40}, 065107 (2013).

\bibitem{Branz:2009yt}
  T.~Branz, T.~Gutsche and V.~E.~Lyubovitskij,
  Phys.\ Rev.\ D {\bf 80}, 054019 (2009).



\bibitem{Lee:2009hy}
  I.~W.~Lee, A.~Faessler, T.~Gutsche and V.~E.~Lyubovitskij,
  Phys.\ Rev.\ D {\bf 80}, 094005 (2009).




\bibitem{Dong:2013iqa}
  Y.~Dong, A.~Faessler, T.~Gutsche and V.~E.~Lyubovitskij,
  arXiv:1306.0824 [hep-ph].

\bibitem{HidalgoDuque:2012ej}
  C.~Hidalgo-Duque, J.~Nieves and M.~P.~Valderrama,
  arXiv:1211.7004 [hep-ph].

\bibitem{review}
  J.~A.~Oller, E.~Oset and A.~Ramos,
  Prog.\ Part.\ Nucl.\ Phys.\  {\bf 45}, 157 (2000).

\bibitem{Voloshin:1976ap} 
  M.~B.~Voloshin and L.~B.~Okun,
  JETP Lett.\  {\bf 23}, 333 (1976)
  [Pisma Zh.\ Eksp.\ Teor.\ Fiz.\  {\bf 23}, 369 (1976)].

\bibitem{Kolomeitsev:2003ac}
  E.~E.~Kolomeitsev and M.~F.~M.~Lutz,
  Phys.\ Lett.\ B {\bf 582}, 39 (2004).

\bibitem{Hofmann:2003je}
  J.~Hofmann and M.~F.~M.~Lutz,
  Nucl.\ Phys.\ A {\bf 733}, 142 (2004).


\bibitem{Guo:2006fu}
  F.~-K.~Guo, P.~-N.~Shen, H.~-C.~Chiang, R.~-G.~Ping and B.~-S.~Zou,
  Phys.\ Lett.\ B {\bf 641}, 278 (2006).

\bibitem{dany}
  D.~Gamermann, E.~Oset, D.~Strottman and M.~J.~Vicente Vacas,
  Phys.\ Rev.\ D {\bf 76}, 074016 (2007).

\bibitem{danyax}
  D.~Gamermann and E.~Oset,
  Eur.\ Phys.\ J.\ A {\bf 33}, 119 (2007).




\bibitem{Faessler:2007gv}
  A.~Faessler, T.~Gutsche, V.~E.~Lyubovitskij and Y.~-L.~Ma,
  Phys.\ Rev.\ D {\bf 76}, 014005 (2007).

\bibitem{Segovia:2008zz}
  J.~Segovia, A.~M.~Yasser, D.~R.~Entem and F.~Fernandez,
  Phys.\ Rev.\ D {\bf 78}, 114033 (2008).

\bibitem{FernandezCarames:2009zz}
  T.~Fernandez-Carames, A.~Valcarce and J.~Vijande,
  Phys.\ Rev.\ Lett.\  {\bf 103}, 222001 (2009).

\bibitem{Gutsche:2010zza}
  T.~Gutsche and V.~E.~Lyubovitskij,
  AIP Conf.\ Proc.\  {\bf 1257}, 385 (2010).


\bibitem{litojuan}
  J.~Nieves and M.~P.~Valderrama,
  Phys.\ Rev.\ D {\bf 86}, 056004 (2012).

\bibitem{HidalgoDuque:2012pq}
  C.~Hidalgo-Duque, J.~Nieves and M.~P.~Valderrama,
  Phys.\  Rev.\ D {\bf 87}, 076006 (2013).



\bibitem{Guo:2008zg}
  F.~-K.~Guo, C.~Hanhart and U.~-G.~Meissner,
  Phys.\ Lett.\ B {\bf 665}, 26 (2008).


\bibitem{Ding:2009vj}
  G.~-J.~Ding, J.~-F.~Liu and M.~-L.~Yan,
  Phys.\ Rev.\ D {\bf 79}, 054005 (2009).


\bibitem{Li:2012ss}
  N.~Li, Z.~-F.~Sun, X.~Liu and S.~-L.~Zhu,
  arXiv:1211.5007 [hep-ph].

\bibitem{hanhart}
  M.~Cleven, Q.~Wang, F.~-K.~Guo, C.~Hanhart, U.~-G.~Meissner and Q.~Zhao,
  Phys.\  Rev.\  {\bf D} 87, 074006 (2013).

\bibitem{Li:2012mqa}
  M.~T.~Li, W.~L.~Wang, Y.~B.~Dong and Z.~Y.~Zhang,
  Int.\ J.\ Mod.\ Phys.\ A {\bf 27}, 1250161 (2012).

\bibitem{Albaladejo:2013aka} 
  M.~Albaladejo, C.~Hidalgo-Duque, J.~Nieves and E.~Oset,
  arXiv:1304.1439 [hep-lat]. To appear in Phys.\ Rev.\ D.


\bibitem{raquelxyz}
  R.~Molina and E.~Oset,
  Phys.\ Rev.\ D {\bf 80}, 114013 (2009).

\bibitem{nicmorus}
  R.~Molina, D.~Nicmorus and E.~Oset,
  Phys.\ Rev.\ D {\bf 78}, 114018 (2008).


\bibitem{hidden1}
  M.~Bando, T.~Kugo, S.~Uehara, K.~Yamawaki and T.~Yanagida,
  Phys.\ Rev.\ Lett.\  {\bf 54}, 1215 (1985).


\bibitem{hidden2}
  M.~Bando, T.~Kugo and K.~Yamawaki,
  Phys.\ Rept.\  {\bf 164}, 217 (1988).

\bibitem{hidden4}
  U.~G.~Meissner,
  Phys.\ Rept.\  {\bf 161}, 213 (1988).
\bibitem{Albaladejo:2008qa} 
  M.~Albaladejo and J.~A.~Oller,
  Phys.\ Rev.\ Lett.\  {\bf 101}, 252002 (2008).
  
\bibitem{gengvec}
  L.~S.~Geng and E.~Oset,
  Phys.\ Rev.\ D {\bf 79}, 074009 (2009).

\bibitem{pdg}
  J.~Beringer {\it et al.}  [Particle Data Group Collaboration],
  Phys.\ Rev.\ D {\bf 86}, 010001 (2012).

\bibitem{migueh1}
  J.~-J.~Xie, M.~Albaladejo and E.~Oset,
  arXiv:1306.6594 [hep-ph].

\bibitem{BESdata} M.Ablikim {\it et al.} [BES Collaboration], Phys. Lett. B ~{\bf 685}, 27 (2010).


\bibitem{polosa}
  G.~Cotugno, R.~Faccini, A.~D.~Polosa and C.~Sabelli,
  Phys.\ Rev.\ Lett.\  {\bf 104}, 132005 (2010).


\bibitem{Shi:2013zn}
  M.~Shi, D.~-L.~Yao and H.~-Q.~Zheng,
  PoS ConfinementX {\bf }, 149 (2012).

\bibitem{Guo:2010tk}
  F.~-K.~Guo, J.~Haidenbauer, C.~Hanhart and U.~-G.~Meissner,
  Phys.\ Rev.\ D {\bf 82}, 094008 (2010).


\bibitem{ollerpsi}
  U.~-G.~Meissner and J.~A.~Oller,
  Nucl.\ Phys.\ A {\bf 679}, 671 (2001).

\bibitem{chiang}
  L.~Roca, J.~E.~Palomar, E.~Oset and H.~C.~Chiang,
  Nucl.\ Phys.\ A {\bf 744}, 127 (2004).

\bibitem{ollerulf}
  J.~A.~Oller and U.~G.~Meissner,
  Phys.\ Lett.\ B {\bf 500}, 263 (2001).

\bibitem{danyjuan}
  D.~Gamermann, J.~Nieves, E.~Oset and E.~Ruiz Arriola,
  Phys.\ Rev.\ D {\bf 81}, 014029 (2010).

\bibitem{weinberg}
  S.~Weinberg,
  Phys.\ Rev.\  {\bf 137}, B672 (1965).

\bibitem{hanhartwei}
  V.~Baru, J.~Haidenbauer, C.~Hanhart, Y.~.Kalashnikova and A.~E.~Kudryavtsev,
  Phys.\ Lett.\ B {\bf 586}, 53 (2004).





\end{thebibliography}

\end{document}